\documentclass[nato]{crckbked}
\numreferences
\usepackage{amssymb,epsf,slashed}
\newcommand{\noi}{\noindent}

\newcommand{\vs}{\vspace}
\newcommand{\hs}{\hspace}
\newcommand{\half}{\frac{1}{2}}

\newcommand{\eq}{\begin{equation}}
\newcommand{\en}{\end{equation}}
\newcommand{\eqa}{\begin{eqnarray}}
\newcommand{\ena}{\end{eqnarray}}

\begin{document}
\thispagestyle{empty}
{
\vspace{-1.0cm}
\scriptsize
\hfill
\parbox{23mm}{
DESY 00-033\\HLRZ2000\_3
             }
}\\[5mm]
{\bf
ON THE ETA-INVARIANT IN THE 4D CHIRAL U(1) 
THEORY}\footnote{Talk by A. Hoferichter at NATO Advanced Research
Workshop
{\sl Lattice Fermions and Structure of 
the Vacuum}, October 1999, Dubna, Russia.}
\\
\\
\\
\noi
\hs*{2.1cm} {V. BORNYAKOV}\\
\hs*{2.1cm} {\it Institute for High Energy Physics IHEP,\\ 
\hs*{2.1cm} 142284 Protvino, Russia}\\ \\
\hs*{2.1cm} {A. HOFERICHTER}\\
\hs*{2.1cm} {\it Deutsches Elektronen-Synchrotron DESY and NIC, \\ 
\hs*{2.1cm} 15735 Zeuthen, Germany} \\ \\
\hs*{2.1cm} {G. SCHIERHOLZ}\\
\hs*{2.1cm} {\it Deutsches Elektronen-Synchrotron DESY, \\ 
\hs*{2.1cm} 22603 Hamburg, Germany \,\,and \,\, \\ 
\hs*{2.1cm} Deutsches Elektronen-Synchrotron DESY and NIC, \\ 
\hs*{2.1cm} 15735 Zeuthen, Germany}\\ \\
\hs*{2.1cm} {A. THIMM}\\
\hs*{2.1cm} {\it Institut f\"ur Theoretische Physik, Freie 
             Universit\"at Berlin, \\ 
\hs*{2.1cm} 14195 Berlin, Germany}\\ 

%\end{opening}
%\begin{document}
\vs*{-0.18cm}
\section{Introduction}
%=====================
In this talk we will focus on the imaginary part of the 
effective action of a four--dimensional chiral U(1) theory. 
Let us recall some
relations in the 
continuum\footnote{We will follow \cite{Alvarez-Gaume:1985xf}.} 
first, as they 
will be our guideline for the 
considerations on the lattice. Consider a compactified Euclidean
space--time of four dimensions with chiral effective action 
$\,W[A]\,$ for a, say, left--handed fermion $\,\psi\,$ formally 
defined through
\eq
{\rm e}^{-W[A]} =
 \int \!\!\! D \bar{\psi} D \psi \,
{\rm e}^{-S [\bar{\psi},\,\psi,\,A]}\: . 
\en
\noi
The fermionic action $\,S[\bar{\psi},\psi,A] \,$ in the presence of 
an external gauge field $\,A_{\mu} = - A_{\mu}^{\dagger}\,$ is given by 
\eq
S[\bar{\psi},\psi,A] = 
\int \!\! d^4 x \:\bar{\psi} \slashed{D}_L [A] \psi \: ,
\en 
\noi 
with the usual definitions of the Dirac operator $\,\slashed{D}\,$ and 
the projectors $\,P_{R,L}$  
\eqa
\slashed{D}_L[A]=\slashed{D}[A] P_L\,,\quad \slashed{D} 
= {\rm i}\gamma^{\mu}(\partial_{\mu} + A_{\mu})\,,\quad
P_{R,L}= \half (1 \pm \gamma_5) \; .
\ena
\noi
To make $\,W[A]\,$ better defined, one applies the doubling trick: 
$\,\slashed{D}_L[A]\,\rightarrow\, {\rm i}\slashed{\partial}P_R 
+ \slashed{D}_L[A]$.
Now, let $\,A_t = (1-t) A_0 + t A\,, \; t\in [0,1],\,$ be a path connecting
some initial configuration $\, A_0\,$ and $\,A\,$ in the same 
topological sector\footnote{We will assume $\,A_0=0$ for definiteness.}. 
Then, it is known by the work of
\cite{Alvarez-Gaume:1985xf} and others that
\eq
{\rm Im} W[A] - {\rm Im} W[A_0] = \pi\eta + 2\pi Q_5(A_t) \,,
\label{main0}
\en
\noi
where $\,\eta=\eta[A]\,$ denotes the eta--invariant \cite{Atiyah} of the 
associated five--dimen\-sional Dirac operator 
$\,\mathbb{H}={\rm i}\gamma_5\partial_t + \slashed{D}[A_t]\,$
and $\,Q_5(A_t)$ is the Chern-Simons form.
While $\,\eta\,$ is gauge invariant, $\,Q_5\,$ encodes the
anomaly 
$\,\delta_{g} W[A] \propto {\rm i}\, \delta_{g}Q_5$, where 
$\,\delta_g\,$ is the variation w.r.t. a gauge transformation $\,g$.
On the other hand, the real part of the effective action 
is basically vector--like.
Denoting the effective action of the associated vector 
theory by $\,W_V$, it is known that
\eq
{\rm Re} W[A] - {\rm Re} W[A_0] 
= \half \Big ( W_{\rm V}[A] - W_{\rm V}[A_0] \Big )\,,
\en
\noi
up to local counterterms.
Thus, in the anomaly free model (i.e., $\,Q_5=0$) the eta--invariant
represents the chiral nature of the theory.
Hence, from the point of view of a practical implementation of a chiral 
gauge theory on the lattice, it is most desirable to have 
the imaginary part of the chiral effective action under control.
Some cases, in which the path integral can be evaluated on the lattice 
are listed below. 
\begin{itemize}
\item[\bf{(a)}] One has analytic knowledge of the 
eta--invariant\footnote{See \cite{aoyama} for a lattice 
definition of $\,\eta\,$ and $\,Q_5\,$ within a five-dimensional approach.} 
as some expression in the gauge fields, which can be managed in
lattice computations.
\item[\bf{(b)}] The eta--invariant is small but not zero: $\eta\,\sim\, 0$.

In this case, one can include the effect of $\,\eta\,$ by re-weighting the
observables $\,{\cal O}\,$ by
$\langle {\cal O} \rangle =\langle\,{\cal O}\,{\rm e}^{i\pi{\eta}}\,\rangle_{Re W}/
\langle\,{\rm e}^{i\pi{\eta}}\,\rangle_{Re W}\,,\:
$
\noi where $\,\langle\: \cdot\: \rangle_{Re W}\,$ means averaging w.r.t.
the vector--like real part of $\,W$, only.  
\item[\bf{(c)}] The eta--invariant vanishes after anomaly cancellation: 
$\eta \,=\,0$.

Consequently, the measure is purely vector--like, no re-weighting 
as in {\bf{(b)}} is necessary.
\end{itemize}
\noi
The initial goal of our investigation is
to determine, which of the cases applies to our  
chiral U(1) model.
\noi
Although there has been enormous progress on the conceptual side of
chiral gauge theories on the lattice, e.g.\cite{chiral_progress}, the 
details, even in the `simplest' four--dimensional model, can be quite involved.
%==========================================
\section{Chiral U(1) Theory on the Lattice}
%==========================================
The lattice fermionic action $\,S_F\,$ under consideration is 
given by\footnote{We use standard lattice notation for one flavor.
$\psi$ denotes a Dirac fermion.}
$$
{S_F}[{\bar \psi}, \psi,U]
 =  %\sum_{x,y}
\,{\bar \psi} \, %_{x}  
{\bf \mathbb{M}}[U]%_{x,y}[U] 
\, \psi \,, %_{y},
$$
\noi 
where $\,U\,$ is an external lattice gauge field with 
(compact) link variables $ U_{\mu}(x) \in \, {\rm U(1)}$. 
The fermion matrix 
$\,
\mathbb{M}[U] = {\slashed{\mathbb{D}}}[U] + \mathbb{W}[U]\,,
$
has Dirac part
\eqa
\slashed{\mathbb{D}}= \slashed{D}P_{R} + \slashed{D}P_{L} &=& 
\left(
\begin{array}{cc}
                   0        &        {D}_{{L}\,\;~} \\
                   {D}_{{R}\;\,~}        &        0 \\

\end{array} \right)
\nonumber
\ena
\noi
and Wilson term

\eqa
\mathbb{W} &=& \left(
\begin{array}{cc}
                   W_{{L}{R}}   &          0            \\
                     0      &        W_{{R}{L}}         \\
\end{array}
\right)   \,.
\nonumber
\ena

\noi
We used the so-called ungauged Wilson term
\eqa
\mathbb{W} &=& -\frac{1}{2}\left(
\begin{array}{cc}
                   \partial^b\partial^f   &          0            \\
                     0      &        \partial^f\partial^b         \\
\end{array}
\right)\,,   
\nonumber
\ena

\noi
where $\partial^{f,b}\,$ denotes the lattice forward and backward
derivative, respectively.
In this case there are no counterterms for the imaginary part 
of the effective action \cite{bodwin}.

\noi
In our four--dimensional chiral model, we have chosen the right-handed component
of the gauge field $U_R$ to be 
trivial, in such a way that $\,{D}_{R} \,\rightarrow\, {\partial}_R $.

\noi
We evaluate the imaginary part of the fermionic effective action,
$$
{\rm Im}\,W = - {\rm Im} \, ( \ln \det \mathbb{M}\,)\,,
$$
\noi
in the continuum fermion approach. For details we
refer to the 
literature (e.g.\cite{bodwin}--\cite{Bornyakov:1998xg}).
The basic idea is to associate bosonic and fermionic
degrees of freedom to different cutoffs 
$\,a^{-1}\,$ and $\,a_f^{-1}\,$, respectively. Here $\,a\,$ denotes the 
lattice spacing of the original lattice, where the original lattice gauge 
field  $\,U^a\,$ resides, and $\,a_f < a \,$ is the lattice spacing of a finer
lattice which `carries' the fermions, in the background of some gauge 
field $\,U^{a_f}\,$ obtained by suitable 
interpolation (e.g. \cite{interpolate}) from $\,U^a$. 
To evaluate the fermionic effective action for the continuum
fermions we have to consider
\eq
- \lim_{a_f\to 0}\,\Big (\, \ln \det \mathbb{M}[U^{a_f}]\,\Big )\,,
\label{eq:W_CFA}
\en
\noi
with $\,a\,$ held fixed. Eventually, also the limit $\,a\rightarrow 0\,$
has to be taken, which we will not perform here, since it is not necessary
for meeting the initial goal.
For perturbative fields, it has been shown that 
(\ref{eq:W_CFA}) exists and has correct properties 
after adding the appropriate counterterms, e.g. \cite{Bornyakov:1998xg}. 
Gauge invariance breaking effects
\footnote{... in the trivial topology sector } 
vanish as $\,O(a_f/a)\,$ 
\cite{bodwin,hernandez}. 

\noi
To compute $\:\det \mathbb{M}[U^{a_f}]\:$ we apply a non-Hermitean 
Lanczos procedure with complete re-orthogonalization. 
In an additional step we construct the anomaly free model by 
imposing the anomaly cancellation 
condition\footnote{
We consider anomaly free models with one right-handed fermion of charge
$\,c_R\,$ and $\,(c_R/c_L)^3\,$ left-handed fermions of 
charge $\,c_L,\:(c_R > c_L > 0)$. For discussion see, e.g. \cite{suzuki}. 
}:
\eq
\sum_{\alpha=1}^{N_f} \epsilon_{\alpha} c_{\alpha}^3 =0\,.
\label{eq:a-cancellation}
\en
\noi
The sum runs over the $\,N_f\,$ different flavors with 
chiralities $\,\epsilon_{\alpha}\,$ and fermion charges $\, c_{\alpha}$. 
In the anomaly free case, we formally define a lattice eta--invariant
by 
$$
\lim_{a_f\to0}\,{\rm Im} W[U^{a_f}] \, \stackrel{!}{=}\, \pi\eta\,;
\quad{a={\rm const}}\,,
$$
\noi
where we have utilized the continuum 
relations\footnote{From now on, we set $a$=1 and
drop the superscript $\,a_f\,$ on interpolated configurations.}.
%====================================
\section{Perturbative Structure}
%====================================
Before turning to numerical results, we will investigate 
the perturbative structure of the imaginary part of the effective action.
%
% fig. one \label{fig:pt0}
%
%
\begin{figure}[hbt]
\begin{center}
\epsfxsize=320pt
\epsfbox{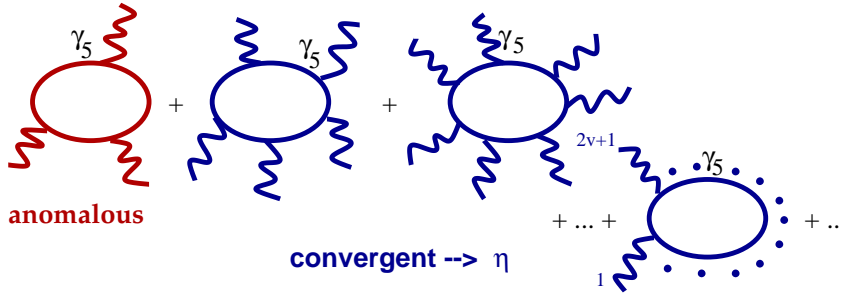}
\caption{Graphical representation of $\,{\rm Im}W[A]$. }
\label{fig:pt0}
\end{center}
\end{figure}
For simplicity, we will stick to continuum relations, as in the introduction.
Fig.\,\ref{fig:pt0} represents $\,{\rm Im} W[A]\,$ in terms of 
graphs, with external field $\,A$.
In our case of four dimensions just the first diagram is 
divergent, while the others are convergent and (up to a factor) 
sum up to $\,\eta[A]$. 
Hence, after anomaly cancellation, the five--leg 
diagram is expected to give the leading contribution 
to $\,{\rm Im} W[A]\,$ for perturbative fields.
Just this diagram would require the 
investigation of $\,O(10000)\,$ terms in the general case.
Therefore, we will choose another 
strategy to probe the perturbative behavior. By using plane-wave 
configurations\footnote{With (small) amplitudes $\,a_{\mu,l},\,b_{\mu,l}\,$ 
and momenta $\,k_{l}$.}
\eq
A_{\mu}(x) = c_{\alpha}\,\sum_{l} \, {a}_{\mu,{l}} \cos (k_{l} x) 
+ {b}_{\mu,l} \sin ( k_{l}x)
\label{eq:planewave}
\en
as numerical input, we rescale the 
charge $\,c_{\alpha}\,\rightarrow\,\xi \, c_{\alpha}$, keeping all 
other parameters 
fixed, and set the corresponding effective actions into relation. 
For small enough amplitudes and charges, we expect for one flavor $\alpha$ 
\eq
\frac{{\rm Im}W_{[\xi\,c_{\alpha}]}}
{{\rm Im}W_{[c_{\alpha}]}} \;\sim\; \xi^{2\nu+1}\:,
\label{eq:scaling}
\en
\noi
where $\,(2 \nu +1)\,$ is the number of external legs of the 
first non-vanishing
diagram in Fig.\,\ref{fig:pt0} and $\,W_{[c_{\alpha}]}\,$ is the effective
action for the given charge $\,c_{\alpha}$. With $\,\xi=2\,$ in
(\ref{eq:scaling}) typical numbers like $\,2^3\,$ or $\,2^5\,$ 
would single out the anomalous or the five--leg diagram as leading 
contribution to $\,{\rm Im}W$, respectively. 
%
% fig. two \label{fig:pt1}
%
\begin{figure}[hbt]
\hs*{0.58cm}\hbox{
\epsfxsize=11.0cm
\epsfbox{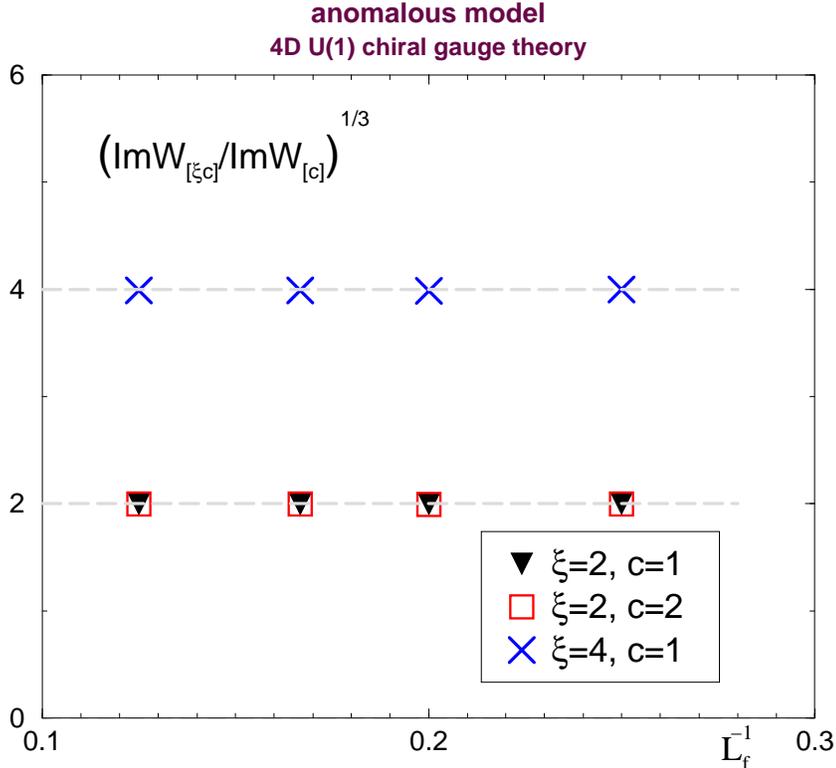}
}
\caption{The $\,a_f$ dependence of (the cubic root of) 
eq.(\ref{eq:scaling}) for different scaling factors $\,\xi$ and 
charges $\,c$. The dashed lines indicate the expectation according to
the anomalous three-leg graph. 
}
\label{fig:pt1}
\end{figure}
\noi Indeed, we find this characteristic behavior as shown 
in Figs.\,\ref{fig:pt1}, \ref{fig:pt2}, where we display the $\,\xi\,$ 
and $\,a_f\,$ dependence of 
eq.(\ref{eq:scaling})\footnote{For convenience, in 
Fig.\,\ref{fig:pt1} we display the cubic root of (\ref{eq:scaling}).}. 
The dashed lines represent the behavior predicted by the anomalous diagram 
(i.e. $\, \nu=1\, $ in Fig.\,\ref{fig:pt1}) and, after anomaly 
cancellation, the five--leg diagram (i.e. $\,\nu=2\,$ in 
Fig.\,\ref{fig:pt2}).
In this way, we nicely reproduce the perturbative structure 
of $\,{\rm Im}W$. 
In general, the imaginary part of the effective action does not 
vanish, but for typical configurations at weak coupling
it might have a small magnitude, such that still the 
case {\bf (b)}, or in some approximation, {\bf (a)} listed 
in the introduction 
is realized.
%
% fig. three \label{fig:pt2}
%
\begin{figure}[hbt]
\hs*{0.58cm}\hbox{
\epsfxsize=11.0cm
\epsfbox{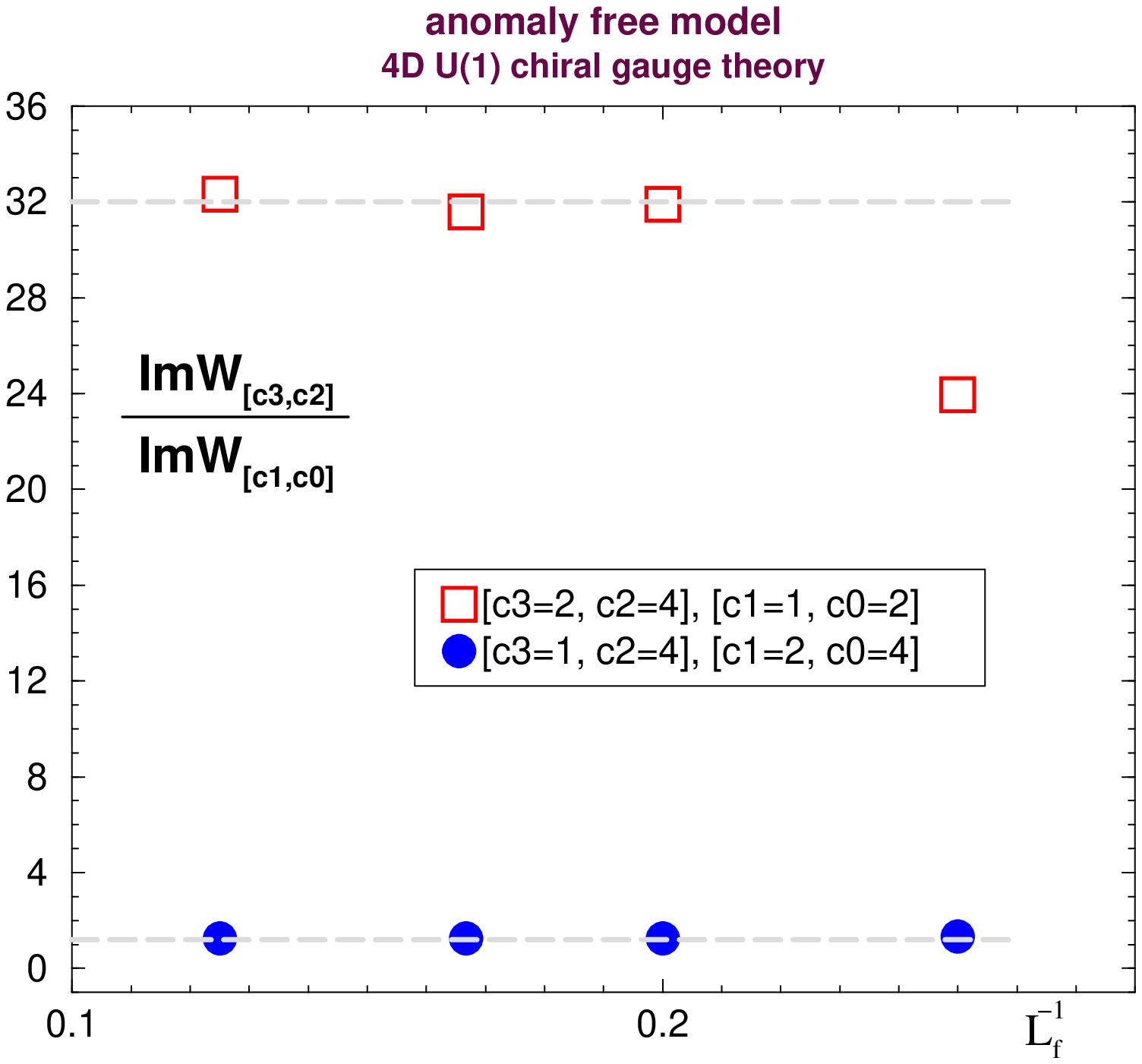}
}
\caption{
The $\,a_f$ dependence of eq.(\ref{eq:scaling}) for the same external 
gauge fields as in Fig.\,\ref{fig:pt1}, but after anomaly cancellation. 
$c_{\rm odd(even)}\,$ denote the charges of the 
left(right)--handed fermions involved, according 
to (\ref{eq:a-cancellation}):  
for instance, $\:[c_1\!=\!1,\,c_0\!=\!2]\,$ is a shorthand notation
for the anomaly free $(11111111\:2)$ model. The dashed lines 
indicate the behavior expected from the five--leg diagram.
 }
\label{fig:pt2}
\end{figure}
%
%==========================
\section{Numerical Results}
%==========================
Here we consider the case of weak coupling 
configurations\footnote{Some results have been 
discussed in \cite{lat99eta}}, without
any topological obstructions (e.g. DeGrand--Toussaint monopoles).
We investigated four gauge configurations, denoted by 
$\,U_{1}, \dots,U_{4}\,$ residing 
on an original lattice of size $\,3^4$ and which were interpolated to
finer lattices up to $\,L_f = 8$.
The interpolation has been refined in order to estimate 
lattice errors. Details have been presented in \cite{axel}.

\noi
The configurations $\,U_1,\, U_2,\, U_3\,$ were generated randomly with 
the constraint that the link 
angles $\,|\,\theta_{\mu}(x)\,| < \pi/5,\,\pi/6,\,\pi/8\,$.\,\,
$U_4\,$ was generated at weak coupling in a (quenched)
vector theory simulation. All original configurations were free of
Dirac plaquettes and monopoles. The plaquette
values on the original lattice are given in Tab.\,\ref{tab:plaq}. 
\begin{center}
\begin{table}[htb]
\begin{tabular}{ccccc}
\hline
  & $U_1$  & $U_2$ & $U_3$ & $U_4$\\
\hline
$1\!-\!\cos\theta_P$ :&  0.251 & 0.169 & 0.098  &0.0062\\
\hline
\end{tabular}
\caption{Plaquette values of the original configurations 
$\,U_1,\dots,\,U_4$.}
\label{tab:plaq}
\end{table}
\end{center}
\vs*{-1.1cm}
%
% fig. four \label{fig:two}
%
\begin{figure}[hbt]
\hs*{0.58cm}
\epsfxsize=11.5cm
\epsfbox{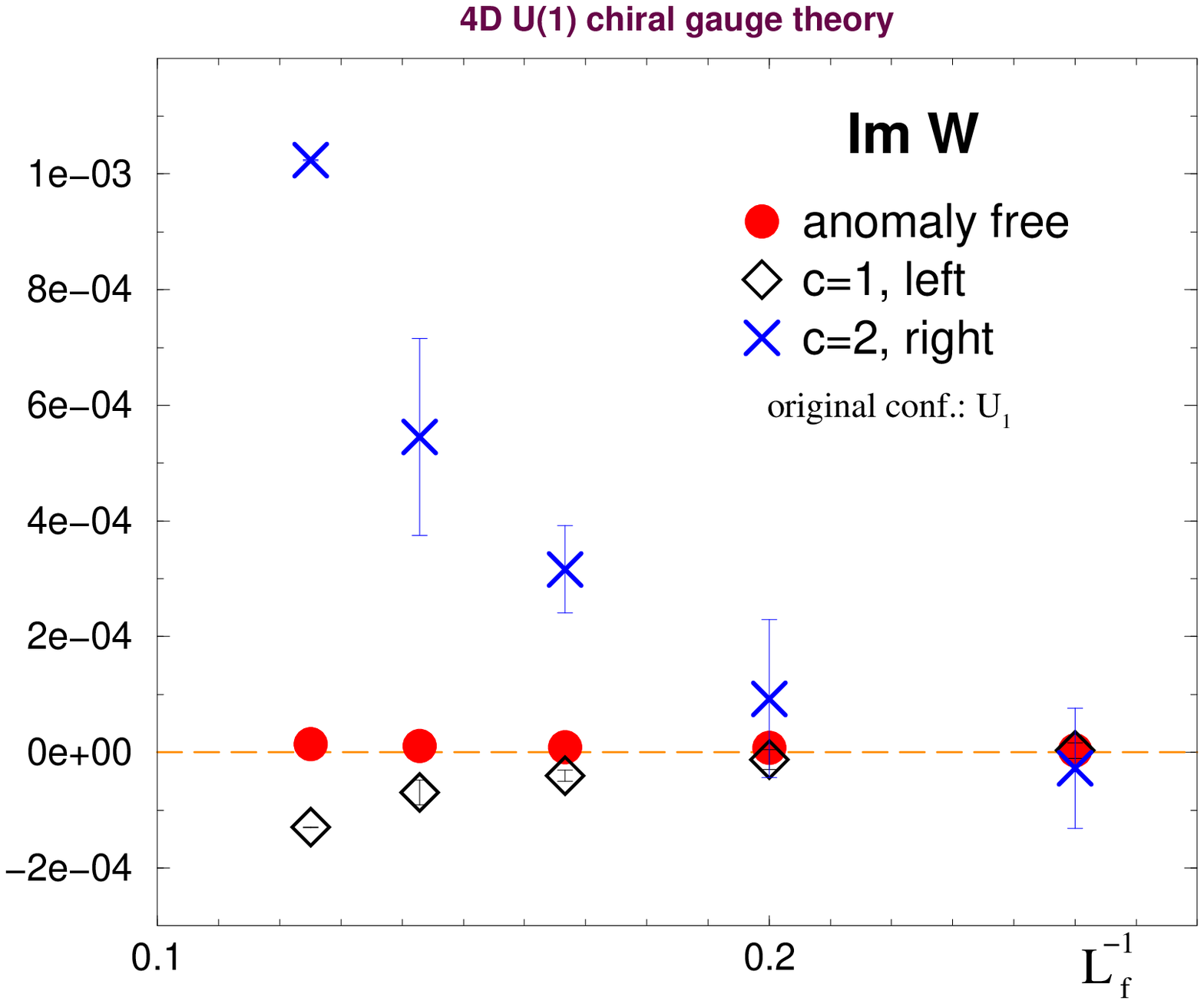}
\caption{
${\rm Im}W\,$ vs. $\,L_f^{-1}\,$ for the 
configuration $\,U_1\,$ with different charges and chiralities.
}
\label{fig:two}
\end{figure}
\noi
In Fig.\,\ref{fig:two} we display the $\,a_f\,$ dependence of $\,{\rm Im} W\,$ 
for the configuration $\,U_1$. The error bars in the figure 
should not be confused
with Monte Carlo errors -- they are an estimation of lattice errors, provided
by the interpolation procedure. 
Since the anomaly is canceled 
event by event, the error bars\footnote{For the finest lattice
we do not display error bars yet.}
can be small in the anomaly free case, although lattice errors
are quite visible in the anomalous charge two--case.
Fig.\,\ref{fig:three} shows the results for the original configuration $\,U_4$.
For configurations $\,U_{2,3}$ we find similar behavior 
and the magnitudes of $\,{\rm Im}W\,$ are in between the cases
of Fig.\,\ref{fig:two} and Fig.\,\ref{fig:three}. 
Despite a clear signal before anomaly 
cancellation, the imaginary part of the effective action 
remains close to zero in the anomaly free model.
We find a difference of up to two orders in magnitude 
for $\,{\rm Im} W\,$ between the models with and without
anomaly. 

\vs*{-0.2cm}
%================
\section{Summary}
%================
In general, the imaginary part of the effective action does not
vanish in the investigated model, as can be shown by perturbative 
analysis.
However, for the given configurations, which we have chosen to 
mimic typical configurations in the weak coupling region, we find 
that $\,{\rm Im} W\,$ basically consists of the 
anomaly. After anomaly cancellation we have at most
$\,|\,{\rm Im}W\,| \,<\, 1.5\cdot 10^{-5}$. 
This implies a very small value of $\,\eta$, even though it 
may take any value in the range 
$[-1,\,1)$.
For weak fields, our investigation favors
options {\bf (a)},\,{\bf (b)} as listed in the introduction. 
Where option {\bf (a)} would involve the evaluation 
of a few (convergent) diagrams of Fig.\,\ref{fig:pt0} by lattice techniques. 
%========================
\section{Acknowledgments}
%========================
This work has been partially supported by the INTAS 96-370 grant.
V.B. acknowledges support from RFBR 99-01230a grant.
The calculations have been partly done on the T3E at ZIB and we thank
H. St\"uben for technical support.
Thanks go to K. Jansen, B. Andreas and K. Scharnhorst for fruitful
discussions.

\noi
We would like to thank all the organizers, in particular V. Mitrjushkin,  
for the warm atmosphere and a wonderful workshop in Dubna. 
%
% Fig. five \label{fig:three}
%
\begin{figure}[hbt]
\hs*{0.58cm}
\epsfxsize=11.5cm
\epsfbox{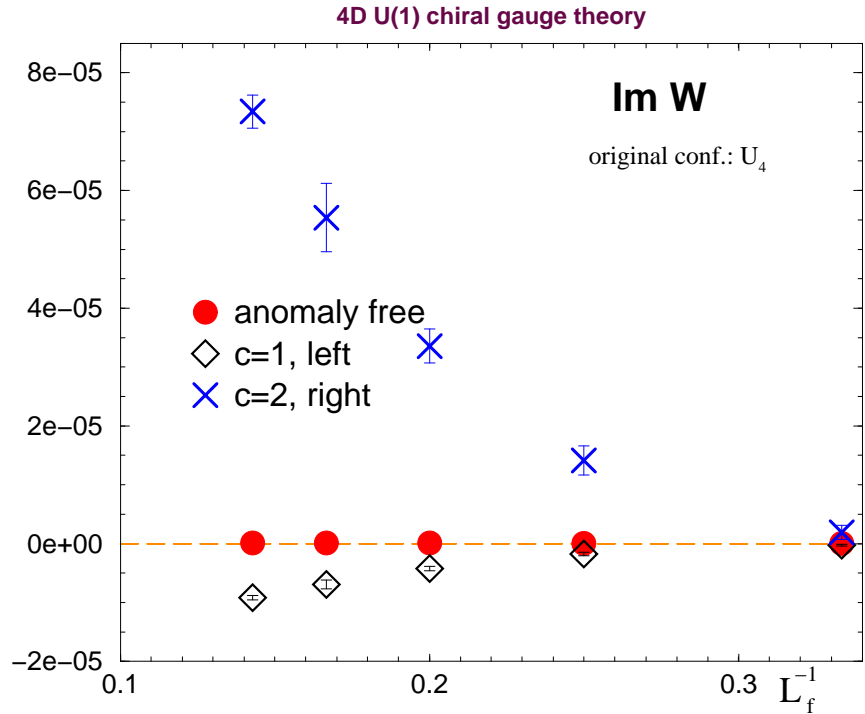}
\caption{
As Fig.\,\ref{fig:two}, but for the
configuration $\,U_4$.
}
\label{fig:three}
\end{figure}
\vs*{-0.18cm}

\end{document}